\documentstyle[epsfig,eqsecnum,aps]{revtex}

\begin{document} 

\draft

\twocolumn[\columnwidth\textwidth\csname@twocolumnfalse\endcsname

\title{Quadrupole deformations of neutron-drip-line nuclei studied within the
Skyrme Hartree-Fock-Bogoliubov approach} \author{M.V.~Stoitsov,$^{1,2}$ J.
Dobaczewski,$^{3}$ P. Ring,$^{2}$ and S.~Pittel$^{4}$}
\address{$^{1}$Institute of Nuclear Research and Nuclear Energy, Bulgarian
Academy of Sciences, Sofia-1784, Bulgaria}
\address{$^{2}$ Physik Department,
Technische Universit\"{a}t, M\"{u}nchen, D-85748 Garching, Germany}
\address{$^{3}$Institute of Theoretical Physics, Warsaw University, Ho\.{z}a 69,
PL-00-681 Warsaw, Poland}
\address{$^{4}$Bartol Research Institute, University of
Delaware, Newark, Delaware 19716}

\maketitle

\begin{abstract}
We introduce a local-scaling point transformation to allow for
modifying the asymptotic properties of the deformed three-dimensional Cartesian
harmonic oscillator wave functions. The resulting single-particle bases  are very
well suited for solving the Hartree-Fock-Bogoliubov equations for deformed
drip-line nuclei.  We then present results of self-consistent calculations
performed for the Mg isotopes and for light nuclei located near the two-neutron
drip line.  The results suggest that for all even-even elements with $Z$=10--18 
the most weakly-bound nucleus has an oblate ground-state shape.
\end{abstract}

\vspace{0.2in}

\pacs{{\bf PACS numbers:} 21.60.Jz, 21.10.Dr, 21.10.Ky}

\addvspace{5mm}]

\narrowtext


\section{Introduction}

Thanks to recent advancess in radioactive ion beam technology, we are now in the
process of exploring the very limits of nuclear binding, namely those regions of
the periodic chart in the neighborhood of the particle drip lines \cite
{[Roe92],[Mue93],[Han93],[Dob98c]}. Several new structure features have already
been uncovered in these studies, including the neutron halo, and others have been
predicted.

In contrast to stable nuclei within or near the valley of beta stability, a
proper theoretical description of weakly-bound systems requires a very careful
treatment of the asymptotic part of the nucleonic density. This is particularly
true in the description of pairing correlations near the neutron drip line, for
which the correct asymptotic properties of quasiparticle wave functions and of
one-particle and pairing densities is essential. In the framework of the
mean-field approach, the best way to achieve such a description is to use the
Hartree-Fock-Bogoliubov (HFB) theory in coordinate-space representation
\cite{[Bul80],[Dob84],[Dob96]}.

Such an approach presents serious difficulties, however, when applied to deformed
nuclei. On the one hand, for finite-range interactions the technical and
numerical problems arising when a two-dimensional mesh of spatial points is used
are so involved that reliable self-consistent calculations in coordinate space
should not be expected soon. On the other hand, for zero-range interactions
existing approaches \cite{[Ter97a],[Taj97]} are able to include only a fairly
limited pairing phase space. The main complication in solving the HFB equations
in coordinate space is that the HFB spectrum is unbounded from below, so that
methods based on a variational search for eigenstates cannot be easily
implemented. Because of this and other difficulties, one has to look for
alternative solutions.

In principle, such an alternative solution is well known in the form of the
configurational representation. In this approach, the system of partial
differential HFB equations are solved by expanding the nucleon quasiparticle wave
functions in an appropriate complete set of single-particle wave functions. In
many applications, an expansion of the HFB wave function in a large harmonic
oscillator (HO) basis of spherical or axial symmetry provides a satisfactory
level of accuracy. For nuclei at the drip lines, however, expansion in an
oscillator basis converges much too slowly to describe the physics of continuum
states \cite{[Dob96]}, which play a critical role in the description of
weakly-bound systems. Oscillator expansions produce wave functions that decrease
too steeply in the asymptotic region at large distances from the center of the
nucleus. As a result, the calculated densities, especially in the pairing
channel, are too small in the outer region and do not reflect correctly the
pairing correlations of such nuclei.

In two recent works \cite{[Sto98a],[Sto98b]}, a new transformed harmonic
oscillator (THO) basis, based on a unitary transformation of the spherical HO
basis, was discussed. This new basis derives from the standard oscillator basis
by a local-scaling point coordinate transformation \cite
{[Sto83],[Sto88a],[Sto91]}, with the precise form dictated by the desired
asymptotic behavior of the densities. The transformation preserves many useful
properties of the HO wave functions. Using the new basis, characteristics of
weakly-bound orbitals for a square-well potential were analyzed and the
ground-state properties of some spherical nuclei were calculated in the framework
of the energy density functional approach\cite{[Sto98b]}. It was demonstrated in
\cite{[Sto98a]} that configurational calculations using the THO basis present a
promising alternative to algorithms that are being developed for coordinate-space
solution of the HFB equations.

In the present work, we develop the THO basis for use in HFB equations of
axially-deformed weakly-bound nuclei. Our main goal here is to present and test
these new theoretical methods. As specific applications, we repeat previous
calculations performed for the chain of Mg isotopes \cite{[Ter97a]}, but for
different effective interactions, and then report a preliminary study of light,
neutron-rich nuclei near the drip line. Extensive calculations throughout the mass
table, together with a more detailed analysis of the pairing interaction, will be
presented in a future publication.

The structure of the paper is the following. The THO basis for deformed nuclei is
introduced in Sec.~\ref{sec2}. In Sec.~\ref{sec3} we present an outline of the
HFB theory and discuss several features of particular relevance to our
investigation. Results of calculations are given in Sec.~\ref{sec4}, and
conclusions are presented in Sec.~\ref{sec5}.


\section{Transformed Harmonic Oscillator Basis}

\label{sec2}

In this section, we introduce a generalized class of local-scaling point
transformations, which in principle act differently in the three Cartesian
directions. Next, we apply this transformation to the three-dimensional Cartesian
HO wave functions and derive the corresponding properties of the local densities.


\subsection{Local-scaling point transformations}

\label{sec2a}

Suppose $\{\varphi_{\alpha}({\bbox{r}})\}$ represents a complete set of
orthonormal single-particle wave functions depending on the spatial coordinate
${\bbox{r}}$. (To simplify the presentation, we suppress the spin and isospin
labels here.) Then, one can introduce a local-scaling point transformation (LST)
of the three dimensional vector space, which is a generalization of the analogous
spherically-symmetric LST \cite{[Sto83],[Sto88a],[Sto91]}, namely

\begin{equation}
\begin{array}{ll}
x\longrightarrow & x^{\prime}\equiv
x^{\prime}(x,y,z)=\frac{x}{{r}}f_x({r}), \\ 
y\longrightarrow &
y^{\prime}\equiv y^{\prime}(x,y,z)=\frac{y}{{r}}f_y({r}), \\
z\longrightarrow & z^{\prime}\equiv z^{\prime}(x,y,z)=\frac{z}{{r}}f_z({r}),
\end{array}
\label{glst}
\end{equation}
where ${r}$=$\sqrt{x^2+y^2+z^2}$. 

The LST
functions $f_k({r})$, $k$=$x$, $y$, or $z$, should have mathematical properties
ensuring that (\ref{glst}) is a valid invertible transformation of the
three-dimensional space. In particular, $f_k({r})$ should be monotonic functions
of ${r}$ such that
\begin{equation}
\begin{array}{lll}
f_k(0)=0 & \mbox{~~and~~}
& f_k(\infty )=\infty \end{array} ,
\label{fbound}
\end{equation}
and should
lead to a non-vanishing Jacobian of the LST (\ref{glst}), i.e.,
\begin{eqnarray}
D &\equiv& \frac{\partial(x^{\prime},y^{\prime},z^{\prime})}{\partial(x,y,z)}
\nonumber \\
&=& \frac{x^2f^{\prime}_xf_yf_z+y^2f_xf^{\prime}_yf_z
+z^2f_xf_yf^{\prime}_z}{{r}^{4}}\neq 0,
\label{jacobi}
\end{eqnarray}
where primes denote derivatives with respect to $r$.

When we apply the LST ( \ref{glst}) to the set of wave functions
$\varphi_{\alpha}({\bbox{r}})$, we obtain another set of single-particle wave functions
\begin{equation}
\psi _{\alpha}(x,y,z)=D^{1/2}\varphi_{\alpha}
\textstyle{
\left(
\frac{x}{{r}} f_x({r}), \frac{y}{{r}}f_y({r}), \frac{z}{{r}}f_z({r}) \right)} .
\label{lstawf}
\end{equation}
Due to the factor $D^{1/2}$ entering
Eq.(\ref{lstawf}), the LST of wave functions is unitary and the new wave
functions $\psi _{\alpha}({\bbox{r}})$ are automatically orthonormal, i.e.,
$\langle\psi_{\alpha}|\psi_{\beta}
\rangle$=$\langle\varphi_{\alpha}|\varphi_{\beta}\rangle$=$
\delta_{\alpha\beta}$.

Summarizing, the LST (\ref{glst}) generates from a given complete set of
orthonormal single-particle wave functions another orthonormal and complete set
of single-particle wave functions (\ref{lstawf}) depending on three
almost-arbitrary scalar LST functions $f_k({r})$. The freedom in the choice of
$f_k({r})$ provides great flexibility in the new set $\{\psi
_{\alpha}({\bbox{r}})\}$, and this opens up the possibility of improving on
undesired properties of the initial set. This is the motivation for the present
study in which we use the LST to modify the incorrect asymptotic properties of
deformed HO wave functions.

\subsection{Transformed harmonic oscillator wave functions}

\label{sec2b}

The anisotropic three-dimensional HO potential with three different oscillator
lengths
\begin{equation}
L_{k} \equiv \frac{1}{b_k}=\sqrt{\frac{\hbar}{m\omega_{k}}},
\label{lengths}
\end{equation}
has the form
\begin{equation}
U({\bbox{r}})=\frac{\hbar^2}{2m}\left(
\frac{x^{2}}{L_{x}^{4}} +\frac{y^{2}}{ L_{y}^{4}}
+\frac{z^{2}}{L_{z}^{4}}\right) .
\label{hop}
\end{equation}

Its eigenstates,
the separable HO single-particle wave functions
\begin{equation}
\varphi_{\alpha}({\bbox{r}})=\varphi _{n_{x}}(x)\varphi _{n_{y}}(y)
\varphi _{n_{z}}(z) ,
\label{2dhowf}
\end{equation}
have a Gaussian asymptotic behavior at large
distances,
\begin{equation}
\varphi _{\alpha}({\bbox{r}\rightarrow \infty })
\sim \exp \left[ 
-\frac{1}{2}
\left(
\frac{x^{2}}{L_{x}^{2}} +\frac{y^{2}}{L_{y}^{2}}
+\frac{z^{2}}{L_{z}^{2}}
\right)
\right] .
\label{hoass}
\end{equation}

Applying the LST (\ref{glst}) to these wave functions leads to the so-called THO
single-particle wave functions (\ref{lstawf}),
\begin{equation}
\psi_{\alpha}({\bbox{r}})=D^{1/2}\textstyle{\ \varphi _{n_{x}}\left( \frac{x
}{{r}}f_x({r)}\right) \varphi _{n_{y}}\left( \frac{y}{{r}}f_y({r)}\right) \varphi
_{n_{z}}\left( \frac{z}{{r}}f_z({r)}\right)} ,
\label{3dtho}
\end{equation}
whose asymptotic behavior is
\begin{equation}
\psi_{\alpha}({\bbox{r}\rightarrow\infty })\sim
\exp \left[ -\frac{1}{2} \left( \frac{x^{2}f^2_x}{L_{x}^{2}r^2}
+\frac{y^{2}f^2_y}{L_{y}^{2}r^2} + \frac{z^{2}f^2_z}{L_{z}^{2}r^2} \right)
\right] .
\label{thoass}
\end{equation}
This suggests that we choose the LST
functions to satisfy the asymptotic conditions
\begin{equation}
f_k(r) = \left\{
\begin{array}{cl}
r & \mbox{~~~for small $r$}, \\[1ex] L_k\sqrt{2\kappa r} &
\mbox{~~~for large $r$}.
\end{array}
\right.
\label{lst1}
\end{equation}
With such a choice, the THO wave functions at small $r$ are identical to the
HO wave
functions (note that with (\ref{lst1}) one obtains $D$=1 at small $r$), while at
large $r$ they have the correct exponential and spherical asymptotic behavior,
\begin{equation}
\psi _{\alpha}({\bbox{r}\rightarrow \infty })\sim e^{-\kappa r}.
\label{thoas}
\end{equation}

\subsection{Parametrization of the LST functions}

\label{sec2c}

In principle, we could use the flexibility of having three different LST
functions $f_k({r})$ and three different oscillator lengths $L_k$ of the original
deformed HO basis to tailor the LST transformation to the shape of the deformed
nucleus under investigation. However, for large HO bases (in the present study we
include HO states up to 20 major shells), the dependence of the total energy on
the basis deformation is very weak, so that minimization of the total energy with
respect to the three oscillator lengths $L_k$ is ill-conditioned  (see discussion
and examples given in Ref.~\cite{[Dob97]}). Therefore, in this study, we use a
spherical HO basis depending on a single common oscillator length $L_0$,
\begin{equation}
L_x=L_y=L_z \equiv L_0=
\frac{1}{b_0}=\sqrt{\frac{\hbar}{m\omega_{0}}}.
\label{length}
\end{equation}
With such a choice, it is natural to set the three LST functions $f_k(r)$ equal
to one another,
\begin{equation}
f_x({r})=f_y({r})=f_z({r}) \equiv f({r}).
\label{lst2}
\end{equation}
This allows us to use exactly the same LST function
$f({r})$ as in the previous studies \cite{[Sto98a],[Sto98b]}. Under conditions
(\ref{lst2}), the Jacobian (\ref{jacobi}) assumes the simpler form
\begin{eqnarray}
D &\equiv&
\frac{\partial(x^{\prime},y^{\prime},z^{\prime})}{\partial(x,y,z)} =
\frac{f^{\prime}({r})f^2({r})}{{r}^{2}}.
\label{jacobi2}
\end{eqnarray}

The parametrization of the LST function $f({r})$ used in
Refs.~\cite{[Sto98a],[Sto98b]} was of the form
\begin{equation}
f({r})=L_0
F\left(\frac{r}{L_0}\right),
\label{lstf}
\end{equation}
with the dimensionless
universal function $F$ of the dimensionless variable ${\cal R}$ defined as
\begin{equation}
F({\cal R})\!=\!\left\{\!\!
\begin{array}{lll}
{\cal R}\left(
1+a{\cal R}^{2}\right) ^{1/3} & \mbox{for} & {\cal R}\leq c\;, \\ \; & \; & \; \\
\sqrt{\frac{d_{-2}}{{\cal R}^{2}} \!+\!\frac{d_{-1}}{{\cal R}} \!+\!d_{0}
\!+\!d_{1}{\cal R} \!+\!d_{L}\ln {\cal R}} & \mbox{for} & {\cal R}>c\;.
\end{array}
\right.
\label{lstf1}
\end{equation}
Two different formulae can be
obtained for the function $F({\cal R})$, one for ${\cal R}\leq c$ and one for
${\cal R}>c$. Imposing the condition that the function should be continuous at
the {\em matching radius} $c$ and that it should have continuous first, second,
third, and fourth derivatives leads to the following requirements for the
constants $d_{-2}$, $d_{-1}$, $d_{0}$, $d_{1}$, and $d_{L}$:
\begin{equation}
\begin{tabular}{l@{}l@{}l}
$d_{-2}$ & = & $\frac{1}{3}{\cal A}c{{^{4}}} (
{243+4050 \gamma +8910 { \gamma ^{2}}+7602 {\gamma ^{3}}+2275 {\gamma ^{4}}} )
,$ \\[1ex]
$d_{-1}$ & = & $-{8}{\cal A}c{{^{3}}} ( {81+1242 \gamma +2745 {\gamma
^{2}} +2340 {\gamma ^{3}}+700 {\gamma ^{4}}} ) ,$ \\[1ex]
$d_{0}$ & = & $2{\cal
A}c^{2}( {1215\gamma +2790 {\gamma }}^{2}{+2415 { \gamma ^{3}}+728 {\gamma
^{4}}} +( 486$ \\[1ex] &  & ${+6480 \gamma +14310 {\gamma ^{2}}+12180 {\gamma
^{3}}+3640 {\gamma ^{4}})\ln c ),}$ \\[1ex]
$d_{1}$ & = & $\frac{8}{3}{\cal A}c (
{{243+2430 \gamma +5211 {\gamma ^{2}} +4380 {\gamma ^{3}}+1300 {\gamma ^{4}}}} )
,$ \\[1ex] $d_{L}$ & = & $-{4 }{\cal A}c^{2} ( {243+3240 \gamma +7155 {\gamma
^{2}} +6090 {\gamma ^{3}}+1820 {\gamma ^{4}}} ) ,$
\end{tabular}
\label{coeflst}
\end{equation}
where ${\gamma}$=$ac^{2}$ and ${\cal A}^{-1}$=${81
{{\left( 1+\gamma \right)} ^{10/3}}}$. In this way, the LST function $f({r})$ is
guaranteed to be very smooth, while still depending on only three parameters,
$L_0$, $a$ and $c$.

{}From (\ref{lstf1}), we see that asymptotically the function $F({\cal
R}$$\rightarrow$$\infty)$$\sim$$\sqrt{d_1 {\cal R}}$. Thus, the LST function
obeys conditions (\ref{lst1}) provided that the parameters satisfy
\begin{equation}
\kappa=\frac{d_1}{2L_0}.
\label{dc}
\end{equation}
Two
different approaches can be used in calculations. One possibility is to minimize
the total energy with respect to $L_0$, $a$, and $c$, obtaining as output the
energetically optimal value of the decay constant $\kappa$. Alternatively,  for a
given choice of $\kappa$, we could eliminate one of the three parameters and
minimize the total energy with respect to the other two. The actual procedure
used in our calculations is described in Sec.~\ref{sec4a}.

\subsection{Axially deformed harmonic oscillator}

\label{sec2d}

In the present study, we restrict our HFB analysis to shapes having axial
symmetry. For this purpose, we use HO wave functions in cylindrical coordinates,
$z$, $\rho$, and $\varphi$,
\begin{equation}
\begin{array}{lll}
x & = & \rho \cos
\varphi \;, \\ y & = & \rho \sin \varphi \;, \\ z & = & z\;,
\end{array}
\label{cyl}
\end{equation}
which allows us to separate the HFB equations into
blocks with good projection $\Omega$ of the angular momentum on the symmetry
axis. [Note that the use of cylindrical coordinates is independent of working
with equal oscillator lengths (\ref{length}).] Since the use of a cylindrical HO
basis is by now a standard technique  (see, e.g.,~Ref.~\cite{[Vau73]}), we give
here only the information pertaining to constructing the cylindrical THO states.

The cylindrical HO basis wave functions are given explicitly by
\begin{equation}
\varphi_{\alpha}(z,\rho,\varphi,s,t)= \varphi_{n_{z}}(z)\varphi_{n_{\rho
}}^{m_{l}}(\rho) \frac{e^{im_{l}\varphi}}{\sqrt{2\pi}}\chi _{m_{s}}(s)\chi
_{m_{t}}(t),
\label{ahowf}
\end{equation}
where the spin $s$ and isospin $t$
degrees of freedom are shown explicitly, $n_{z}$ and $n_{\rho}$ are the number of
nodes along $z$ and $\rho$ directions, respectively, while $m_{l}$ and $m_{s}$
are the components of the orbital angular momentum and the spin along the
symmetry axis. The only conserved quantum numbers in this case are the total
angular momentum projection $ \Omega$=$m_{l}$+$m_{s}$ and the parity
$\pi$=$(-)^{n_{z}+m_{l}}$.

In the axially deformed case, the general LST (\ref{glst}) acts only on the
cylindrical coordinates $z$ and $\rho $ and takes the form
\begin{equation}
\begin{array}{llll}
\rho & \longrightarrow &
\rho^{\prime}\equiv\rho^{\prime}(\rho,z) & = \frac{ \rho }{{r}}f_\rho({r}),
\\[1ex]
z & \longrightarrow & z^{\prime}\equiv z^{\prime}(\rho,z) & =
\frac{z}{{r}} f_z({r}),
\end{array}
\label{clst}
\end{equation}
with the
corresponding Jacobian given by
\begin{eqnarray}
D &\equiv&
\frac{\partial(x^{\prime},y^{\prime},z^{\prime})}{\partial(x,y,z)} =
\frac{\rho^2f^{\prime}_\rho f_\rho f_z + z^2f^2_\rho f^{\prime}_z}{{r}^{4}} .
\label{jacobi3}
\end{eqnarray}
Finally, the axial THO wave functions are
\begin{eqnarray}
\psi_{\alpha}(z,\rho,\varphi,s,t) &=& D^{1/2}\textstyle{\
\varphi_{n_{z}}\left(\frac{z}{{r}}f_z({r})\right) \varphi_{n_{\rho
}}^{m_{l}}\left(\frac{\rho }{{r}}f_\rho({r})\right)}  \nonumber \\
&\times&
\frac{e^{im_{l}\varphi}}{\sqrt{2\pi}}\chi _{m_{s}}(s)\chi _{m_{t}}(t).
\label{atho}
\end{eqnarray}

The assumption of a single oscillator length (see Sect. \ref{sec2c}) that we make
in our calculations translates in the axial case to

\begin{equation} L_\rho=L_z \equiv
L_0=\frac{1}{b_0}=\sqrt{\frac{\hbar}{m\omega_{0}}}, \label{lengthax}
\end{equation} \begin{equation} f_\rho({r})=f_z({r}) \equiv f({r}),  \label{lst3}
\end{equation} and the Jacobian (\ref{jacobi3}) reduces to expression
(\ref{jacobi2}).

\subsection{THO and Gauss integration formulae}

\label{sec2e}

At first glance, the THO wave functions (\ref{3dtho}) and (\ref{atho}) look much
more complicated than their  HO counterparts (\ref{2dhowf}) and (\ref{ahowf}). In
particular, in contrast to the HO wave functions, the THO wave functions are not
separable either in the $x$, $y$, and $z$ Cartesian coordinates or in the $\rho$
and $z$ axial coordinates. Due to the presence of the Jacobian factor and the
$r$-dependence of the LST functions, the local-scaling transformation mixes the
$x$, $y$ and $z$ coordinates and the $\rho$ and $z$ coordinates. Nevertheless, as
we now proceed to show, the THO wave functions are readily tractable in any
configurational self-consistent calculation. Indeed, the modifications required
to transform a code from the HO to the THO basis are minor.

One of the properties of the HO basis that makes it so useful is the high
accuracy that can be achieved when calculating matrix elements using
Gauss-Hermite and/or Gauss-Laguerre integration formulae \cite{[Abr70]}. This
feature has been exploited frequently in various mean-field nuclear structure
calculations (see, e.g.,~Refs.~\cite {[Vau73],[Gam90],[Dob97]}). To illustrate
how the same methods can be applied in the THO basis, we focus on the specific
example of a diagonal matrix element of a spin and isospin independent potential
function $V$. This matrix element  can be expressed in the axial HO
representation as
\begin{equation}
\langle\varphi_{\alpha }|V|\varphi_{\alpha
}\rangle =\int\limits_{-\infty }^{\infty}dz\int\limits_{0}^{\infty } \rho \;d\rho
V(z,\rho) \varphi^2_{n_{z}}(z){\varphi_{n_{\rho }}^{m_{l}}}^2(\rho),
\label{matrixho}
\end{equation}
and in the THO representation as
\begin{eqnarray}
\langle\psi_{\alpha }|V|\psi_{\alpha }\rangle &=&\int\limits_{-\infty
}^{\infty}dz\int\limits_{0}^{\infty } \rho \;d\rho V(z,\rho)  \nonumber \\
&\times& D(z,\rho) \textstyle{\ \varphi^2_{n_{z}}\left(\frac{z}{{r}}f_z({r}
)\right) {\varphi_{n_{\rho }}^{m_{l}}}^2\left(\frac{\rho }{{r}}f_\rho({r}
)\right)}.
\label{matrixtho2}
\end{eqnarray}
The way to calculate the second
matrix element (\ref{matrixtho2}) is by first transforming to the $\rho^{\prime}$
and $z^{\prime}$ variables (\ref{clst}). This absorbs the Jacobian $D(z,\rho)$
and leads to an integral over HO wave functions that is almost identical to
(\ref{matrixho}), namely
\begin{eqnarray}
\langle\psi_{\alpha }|V|\psi_{\alpha
}\rangle &=&\int\limits_{-\infty }^{\infty}dz^{\prime}\int\limits_{0}^{\infty }
\rho^{\prime}\;d\rho^{\prime}V\left(z(z^{\prime},\rho^{\prime}),\rho(z^{\prime},
\rho^{\prime})\right)  \nonumber \\
&\times&
\varphi^2_{n_{z}}(z^{\prime}) {\varphi_{n_{\rho }}^{m_{l}}} ^2(\rho^{\prime}).
\label{matrixtho3}
\end{eqnarray}

The only complication in numerically carrying out the integral (\ref{matrixtho3})
involves determining the inverse LST transformations
$z$=$z(z^{\prime},\rho^{\prime})$ and $\rho$=$ \rho(z^{\prime},\rho^{\prime})$
to be inserted into the known function $V(z,\rho)$. But this only has to be done
once, and, moreover, if Gauss quadratures are used to evaluate the integrals, the
inverse transformation only has to be known  at a finite number of
Gauss-quadrature nodes.

Generalization of the above approach to include differential operators, as will
often arise in THO basis configurational calculations, is fairly straightforward.
Such integrals can be done by first transforming derivatives $\partial/\partial
z$ and $\partial/\partial\rho$ into derivatives $\partial/\partial z^{\prime}$
and $\partial/\partial\rho^{ \prime}$, and then performing the integrations in
the variables $z^{\prime}$ and $ \rho^{\prime}$ over ordinary HO wave functions
(see the next section).

\subsection{THO and local densities}

\label{sec2f}

In calculations using the Skyrme force, or in any other calculation that relies
on the local density approximation, we can simplify the THO methodology of
Sec.~\ref {sec2e} even further. Indeed, suppose the mean-field calculation in
question relies on knowing the density matrix $\rho _{\alpha \alpha ^{\prime }}$
in the THO basis. Then the spatial nonlocal density can be expressed as
\begin{equation}
\rho (\bbox{r}_{1},\bbox{r}_{2})={\displaystyle\sum_{\alpha
\alpha ^{\prime }}}\psi _{\alpha }\left( \bbox{r}_{1}\right) \rho _{\alpha \alpha
^{\prime }}\psi _{\alpha ^{\prime }}^{*}\left( \bbox{r}_{2}\right) ,
\label{thoden}
\end{equation}
and the corresponding standard local densities
\cite{[Eng75]} as
\begin{mathletters}
\label{densit}
\begin{eqnarray}
\rho(\bbox{r}) &=&\rho (\bbox{r},\bbox{r})  \label{densit-a} \\ \tau (\bbox{r})
&=&\sum_{k=x,y,z}\left[ \nabla _{k}^{(1)}\nabla _{k}^{(2)}\rho
(\bbox{r}_{1},\bbox{r}_{2})\right] _{\bbox{r}_{1}=\bbox{r} _{2}}
\label{densit-b} \\ j_{k}(\bbox{r}) &=&\frac{1}{2i}\left[ \left( \nabla
_{k}^{(1)}-\nabla _{k}^{(2)}\right) \rho (\bbox{r}_{1},\bbox{r}_{2})\right]
_{\bbox{r}_{1}= \bbox{r}_{2}}
\label{densit-c}
\end{eqnarray}
where
\end{mathletters}
\begin{equation}
\nabla _{k}^{(i)}=\frac{\partial }{\partial
(\bbox{r}_{i})_{k}},
\label{nabla}
\end{equation}
for $i$=1 or 2, and $k$=$x$,
$y$, or $z$. To simplify the notation in Eq.~(\ref {thoden}), we have neglected
the spin and isospin degrees of freedom and, consequently, have shown only the
spin-independent densities (\ref{densit}). Analogous formulae for the spin-dependent
densities $\bbox{s}$, $\bbox{T}$, and $J_{kl}$ \cite{[Eng75]} are
straightforward.

A direct calculation of the derivatives in Eqs.~(\ref{densit}) [after inserting the THO
wave functions (\ref{3dtho}) or (\ref{atho}) into the nonlocal density matrix
(\ref{thoden})] is prohibitively difficult. Fortunately, nothing of the sort is
necessary. It is enough to note that the densities (\ref{densit}) serve almost uniquely
to define the central, spin-orbit, and effective-mass terms of the mean-field
Hamiltonian (see, e.g.,~Refs.~\cite {[Eng75],[Dob97]}), and that these terms are
in turn used to calculate matrix elements through integrals of the type
(\ref{matrixtho3}). Therefore, the densities (\ref{densit}) have to be effectively known
only at selected points $x^{\prime }$, $y^{\prime }$, $z^{\prime }$ (the
Gauss-quadrature nodes) of the inverse LST.

Towards this end, we insert the THO wave functions into the nonlocal density
(\ref{thoden}), which gives
\begin{equation}
\rho(\bbox{r}_{1},\bbox{r}_{2})=D^{1/2}(\bbox{r}_{1})D^{1/2}(\bbox{r} _{2})\rho
^{\prime }\Big(\bbox{r}_{1}^{\prime }(\bbox{r}_{1}),\bbox{r} _{2}^{\prime
}(\bbox{r}_{2})\Big)  ,
\label{hoden1}
\end{equation}
with
\begin{equation} \rho^{\prime }(\bbox{r}_{1}^{\prime },\bbox{r}_{2}^{\prime })
={\displaystyle
\sum_{\alpha \alpha ^{\prime }}}\varphi _{\alpha }\left( \bbox{r} _{1}^{\prime
}\right) \rho _{\alpha \alpha ^{\prime }}\varphi _{\alpha ^{\prime }}^{*}\left(
\bbox{r}_{2}^{\prime }\right) .
\label{hoden2}
\end{equation}
The density matrix
$\rho ^{\prime }(\bbox{r}_{1}^{\prime },\bbox{r} _{2}^{\prime })$ is a standard
object expressed in terms of ordinary HO wave functions, and it can be calculated
using methods that are employed in any code that works in the HO basis. Likewise,
the corresponding local densities
\begin{mathletters}
\label{densitp}
\begin{eqnarray}
\rho ^{\prime }(\bbox{r}^{\prime }) &=&\rho ^{\prime
}(\bbox{r}^{\prime }, \bbox{r}^{\prime })
\label{densitp-a}
\\
\tau_{km}^{\prime }(\bbox{r}^{\prime })
&=&\left[ \nabla _{k}^{(1)^{\prime }}\nabla
_{m}^{(2)^{\prime }}\rho ^{\prime }(\bbox{r}_{1}^{\prime },\bbox{r} _{2}^{\prime
})\right] _{\bbox{r}_{1}^{\prime }=\bbox{r}_{2}^{\prime }}
\label{densitp-b}
\\
j_{k}^{\prime }(\bbox{r}^{\prime }) &=&\frac{1}{2i}\left[ \left( \nabla
_{k}^{(1)^{\prime }}-\nabla _{k}^{(2)^{\prime }}\right) \rho ^{\prime }(
\bbox{r}_{1}^{\prime },\bbox{r}_{2}^{\prime })\right] _{\bbox{r}_{1}^{\prime
}=\bbox{r}_{2}^{\prime }}
\label{densitp-c}
\end{eqnarray}
can be calculated
without any reference to the THO basis. The only complication is that now we have
to calculate the complete kinetic energy tensor density $\tau _{km}^{\prime }$
(\ref{densitp-b}), while finally only its trace (\ref{densit-b}) is needed.
Inserting expression (\ref{hoden1}) into (\ref{densit}), and expressing the differential
operators (\ref{nabla}) as
\end{mathletters}
\begin{equation}
\nabla_{k}^{(i)}=\sum_{m=x,y,z}D_{k}^{m}\nabla _{m}^{(i)^{\prime }},
\label{nabla2}
\end{equation}
for
\begin{equation}
D_{k}^{m}\equiv \frac{\partial
\bbox{r}_{m}^{\prime }}{\partial \bbox{r}_{k}} =\frac{f_{m}}{r}\delta
_{mk}+\frac{rf_{m}^{\prime }-f_{m}}{r^{3}}\bbox{r}_{m} \bbox{r}_{k},
\label{nabla3}
\end{equation}
we obtain that
\begin{mathletters}
\label{densitf}
\begin{eqnarray}
\rho (\bbox{r}(\bbox{r}')) &=&D\rho ^{\prime }(\bbox{r}^{\prime
}),
\label{densitf-a}
\\
\tau (\bbox{r}(\bbox{r}'))
&=&D{\displaystyle\sum_{kmn}}D_{n}^{k}D_{n}^{m} \tau _{km}^{\prime
}(\bbox{r}^{\prime })  \nonumber \\
&&+\frac{1}{2}{\displaystyle\sum_{km}}\left[
\nabla _{k}D\right] D_{k}^{m}\left[ \nabla _{m}^{\prime }\rho ^{\prime
}(\bbox{r}^{\prime })\right]   \nonumber \\
&&+\frac{1}{4}D^{-1}\left[
\bbox{\nabla}D\right] ^{2}\rho (\bbox{r}^{\prime }),
\label{densitf-b}
\\
j_{k}(\bbox{r}(\bbox{r}')) &=&D{\displaystyle\sum_{m}}D_{k}^{m}j_{m}^{\prime
}(\bbox{r}^{\prime }).
\label{densitf-c}
\end{eqnarray}

To use formulae (\ref{densitf}), we must calculate the Jacobi matrix $D_{k}^{m}$ and its
determinant $D$ at points $\bbox{r}(\bbox{r}')$; however, this need be done only
once for all iterations. On the other hand, no inverse LST needs to be performed
for the densities, because expressions (\ref{densitf}) give directly the values of the
local densities at the inverse LST points, as required in matrix-element
integrals of the type (\ref{matrixtho3}).

\section{Hartree-Fock-Bogoliubov theory}

\label{sec3}

Hartree-Fock-Bogoliubov (HFB) theory \cite{[RS80]} is based on the Ritz
variational principle applied to the many-fermion Hamiltonian,
\end{mathletters}
\begin{equation}
H=\sum_{\alpha\alpha^{\prime }}t_{\alpha\alpha^{\prime
}}a_{\alpha}^{\dagger }a_{\alpha^{\prime }}+\sum_{\alpha\alpha^{\prime
}\beta\beta^{\prime }}\bar{v }_{\alpha\alpha^{\prime }\beta\beta^{\prime
}}a_{\alpha}^{\dagger }a_{\alpha^{\prime }}^{\dagger }a_{\beta^{\prime
}}a_{\beta},
\label{ham}
\end{equation}
with trial functions in the form of a
quasiparticle vacuum. The resulting HFB equations can be written in matrix form
as
\begin{equation}
\left(
\begin{array}{cc}
h - \lambda & \Delta \\ -\Delta ^{*}
& -h^{*}+\lambda \end{array} \right) \left( \begin{array}{l} U_{n} \\ V_{n}
\end{array}
\right)
=E_{n}
\left(
\begin{array}{l}
U_{n} \\
V_{n}
\end{array}
\right) \;,
\label{hfbeq}
\end{equation}
where $E_{n}$ are the quasiparticle
energies, $\lambda$ is the chemical potential, and the matrices $h=t+\Gamma$ and
$\Delta$ are defined by the matrix elements of the two-body interaction
\begin{equation}
\begin{array}{rcl}
\Gamma _{\alpha\alpha^{\prime }} & = &
\sum\limits_{\beta\beta^{\prime }}
\bar{v}_{\alpha\beta\alpha^{\prime}\beta^{\prime }} \rho _{\beta^{\prime }\beta},
\\
\Delta _{\alpha\alpha^{\prime}} & = & {\textstyle\frac{1}{2}}
\sum\limits_{\beta\beta^{\prime }} \bar{v}_{\alpha\alpha^{\prime
}\beta\beta^{\prime }} \kappa_{\beta\beta^{\prime}},
\end{array}
\label{rhokm}
\end{equation}
$\rho _{\beta^{\prime }\beta}$ and $\kappa_{\beta\beta^{\prime}}$
being the density matrix and pairing tensor, respectively. HFB theory is by now a
standard tool in nuclear structure calculations, and we refer the reader to
Ref.~\cite{[RS80]} for details. Below we discuss several features of the
formalism that are especially pertinent to the present application, namely
canonical states, the pairing phase space, and those quantities that dictate the
stability of a nucleus with respect to two-neutron emission.

\subsection{Canonical states}

\label{sec3a}

Canonical states are defined as the states that diagonalize the HFB one-body
density matrix $\rho({\bbox{r}_1},{\bbox{r}_2})$ of Eq.~(\ref{thoden}), i.e.,
\begin{equation}
\int \rho({\bbox{r}_1},{\bbox{r}_2})\breve\psi_{i}({\bbox{r}_2})d{\bbox{r}_2 }
=v_{i}^{2}\breve\psi_{i}({\bbox{r}}_1)\;,
\label{can1}
\end{equation}
where, due
to the Pauli principle, the canonical occupation numbers $ v_{i}^{2}$ obey the
condition $0\leq v_{i}^{2}\leq 1$.

{}For self-consistent solutions, the canonical occupation numbers $v_{i}^{2}$ are
determined by the diagonal matrix elements $h_{ii}$ and $\Delta _{i\bar{i }}$ of
the particle-hole (p-h) and particle-particle (p-p) Hamiltonians in the canonical
basis via the following BCS-like equation \cite{[RS80]}:
\begin{equation}
v_{i}^{2}=\frac{1}{2}-\frac{h_{ii}-\lambda}{2E_{i}},
\label{bcsuv}
\end{equation}
where
\begin{equation}
E_{i}=\sqrt{\left(h_{ii}-\lambda
\right)^{2}+\Delta_{i\bar{i}}^{2}} ~.
\label{bcsqpe}
\end{equation}
The chemical
potential $\lambda$ is determined from the particle number condition
\begin{equation}
N=\sum\limits_{i}\;v_{i}^{2}=\sum_n N_n,
\label{bcsn}
\end{equation}
where $N_n$ denote the norms of the lower HFB wave functions of
Eq.~(\ref{hfbeq}), i.e.,
\begin{equation}
N_n = \sum_\alpha V^2_{\alpha n} ~.
\end{equation}
In the canonical representation, the average (proton or neutron)
pairing gap $\widetilde{\Delta}$ \cite{[Dob84]} is given by the average value of
$\Delta _{i\bar{i}}$ in the corresponding (proton or neutron) canonical states,
\begin{equation}
\widetilde{\Delta}=\frac{1}{N}\sum\limits_{i}\;
\Delta_{i\bar{i}}v_{i}^{2}~,
\label{avdel}
\end{equation}
where $N$ is the number of nucleons of that type (see Eq.~~(\ref{bcsn})).

Whenever infinite complete single-particle bases are used in configurational
calculations, one may freely expand the upper and lower HFB wave functions of
Eq.~(\ref{hfbeq}) (the quasiparticle wave functions), as well as the standard
eigenstates of the p-h Hamiltonian $h$, in the canonical basis. These expansions
are often extremely slowly converging, however, and any truncation of the basis
typically induces large errors. Therefore, in practice, when working with finite
bases, one should not expand quasiparticle wave functions and the single-particle
eigenstates of $h$ in the canonical basis. The reason is very simple, and it
stems from different asymptotic properties of these objects. As discussed in
Ref.~\cite{[Dob84]}, the quasiparticle spectrum and wave functions are partly
discrete and localized and partly continuous and asymptotically oscillating,
respectively. These properties are completely analogous to properties of the
eigenstates of $h$, which are also discrete and localized (for negative
eigenenergies) or continuous and oscillating (for positive eigenenergies). On the
other hand, the properties of eigenvalues and eigenstates of the density matrix
(\ref{can1}) are very different, namely the entire spectrum is discrete and all
the wave functions are localized. Therefore, even if formally the set of
canonical states is complete, it is extremely difficult to expand any oscillating
wave function in this basis.

These considerations make it clear that the optimum way of solving the HFB
equations is by using the coordinate representation, in which the various
asymptotic properties are in a natural way correctly fulfilled. This technique is
widely used when spherical symmetry is imposed; then one only has to solve
systems of one-dimensional differential equations, which is an easy task. On the
other hand, the case of axial symmetry requires solving two-dimensional
equations, and that of triaxial shapes requires working with a three-dimensional
problem. None of these two latter cases has up to now been effectively solved in
coordinate space, although work on the axial solutions is in progress
\cite{[Obe99]}.

Therefore, without having access to coordinate-representation solutions, we are
obliged to use methods based on a configurational expansion. In this respect, one
may clearly distinguish two classes of finite single-particle bases, each of
which aims at a reasonable solution of the HFB equations (\ref{hfbeq}). One uses
a truncated basis composed of eigenstates of $h$ \cite
{[Ter96],[Ter97a],[Taj97]}. This basis is partly composed of discrete localized
states and partly of discretized continuum and oscillating states. Technically it
is very difficult to include many continuum states in the basis, especially when
triaxial deformations are allowed. In practice,
Refs.\cite{[Ter96],[Ter97a],[Taj97]} included states up to several MeV into the
continuum. Such a small phase space is certainly insufficient to describe spatial
properties of nuclear densities at large distances, although some ground-state
properties, like total binding energies, will be at most weakly affected.

The second uses a truncated infinite discrete basis. The most common of course is
the HO basis, which has been used in numerous HFB calculations, especially those
employing the Gogny effective interaction (see, e.g.,~Refs.~\cite
{[Gog75],[Gir83],[Egi80],[Egi95]}), and in Hartree-Bogoliubov calculations based
on a relativistic Lagrangian (see, e.g.,~Refs.~\cite {[Afa96],[Rin96]}). Because
it uses a basis with a similar structure to the canonical basis (infinite and
discrete), this approach can be viewed as aiming at the best possible
approximation to the canonical states and not the quasiparticle states. In this
sense, the amplitudes $U_n$ and $V_n$ that appear in Eq.~(\ref{hfbeq}) should be
considered more as expansion coefficients of quasiparticle states in a basis
similar to the canonical basis than as quasiparticle wave functions themselves.

Our approach, which we discuss in greater detail below, belongs to the second
class. The THO basis defined and described in Sec.~\ref{sec2} is a model that
aims at an optimal description of the canonical states. Therefore, in the
following we adapt properties of the THO basis, and in particular the value of
the decay constant $\kappa$ (\ref{thoas}), to the asymptotic properties of
canonical states. In fact, the unique decay constant of all THO basis states is
exactly the desired property of canonical states. As discussed in
Ref.~\cite{[Dob96]}, the asymptotic properties of the most important canonical
states (those having average energies close to the Fermi energy) are governed by
a common unique decay constant,
\begin{equation}
\kappa=\sqrt{\frac{2m(E_{\min}-\lambda)}{\hbar^2}} ~,
\label{hfbdc}
\end{equation}
where $E_{\min}$ is the lowest quasiparticle energy $E_n$. This
should be contrasted with decay constants associated with the eigenstates of $h$,
which are all different and depend on the single-particle eigenenergies.

\subsection{The cut-off procedure}

\label{sec3b}

HFB calculations in configurational representation invariably require a
truncation of the single-particle basis and a truncation in the number of
quasiparticle states. The latter is usually realized by defining a cut-off
quasiparticle energy $E_{\max}$ and then including quasiparticle states only up
to this value. When the finite-range Gogny force is used both in the p-p and p-h
channels, the cut-off energy $E_{\max }$ has numerical significance only. In
contrast, HFB calculations based on Skyrme forces in the p-h and p-p channels, as
well as any other calculations based on a zero-range force in the p-p channel
\begin{equation} 
V^{\delta }({\bbox{r}},{\bbox{r}}^{\prime })=V_{0}\delta
({\bbox{r}}-{ \bbox{r}}^{\prime })\;  
\label{delta} 
\end{equation} 
{\em require}
a finite space of states. This is because, for any value of the coupling constant
$V_0$, they give divergent energies with increasing $E_{\max}$ (see the
discussion in Ref.~\cite{[Dob96]}).

The choice of an  appropriate cut-off procedure has been discussed in the case of
coordinate-space HFB\ calculations for spherical nuclei \cite{[Dob84]}. It was
demonstrated there that one must sum up contributions from all states close in
quasiparticle energy to the bound particle states to obtain correct density
matrices in the HFB method. Since the bound particle states are associated with
quasiparticle energies smaller than the absolute value $D$ of the depth of the
effective potential well, one had to take the cut-off energy $E_{\max}$
comparable to $D$.

In the case of deformed HFB calculations, and especially when performing
configurational HFB calculations, it is difficult to look for the depth of the
effective potential well in each $\Omega^{\pi }$ subspace. Thus, an alternative
criterion with respect to the above cut-off procedure used in spherical
calculations is needed. For this purpose, we have adopted the following procedure
(see Appendix B of \cite{[Dob84]}). After each iteration, performed with a given
chemical potential $\lambda$, we calculate an auxiliary spectrum $\bar{e}_{n} $
and pairing gaps $\bar{\Delta}_{n}$ by using for each quasiparticle state the
BCS-like formulae,
\begin{mathletters}
\begin{eqnarray}
E_{n} &=&\sqrt{\left(\bar{e}_{n}-\lambda\right)^{2}+\bar{\Delta}_{n}^{2}}, \\
N_{n} &=&\frac{1}{2}-\frac{\bar{e}_{n}-\lambda}{2E_n},
\label{equi2}
\end{eqnarray}
or equivalently
\end{mathletters}
\begin{mathletters}
\begin{eqnarray}
\bar{e}_{n} &=& (1-2N_n)E_n,
\label{equi3} \\
\bar{\Delta}_{n}&=& 2E_n\sqrt{N_n(1-N_n)}.
\label{equi4}
\end{eqnarray}
Then, in the next
iteration, we readjust the proton and neutron chemical potentials to obtain the
correct values of the proton and neutron particle numbers (\ref{bcsn}), where
again $N_n$ is calculated for the equivalent spectrum, Eq.~(\ref{equi2}). Due to
the similarity between the equivalent spectrum $\bar{e}_{n}$ and the
single-particle energies, we are taking into account only those quasiparticle
states for which
\end{mathletters}
\begin{equation}
\bar{e}_{n}\leq\bar{e}_{\max},
\label{cutoff}
\end{equation}
where
$\bar{e}_{\max}$$>$0 is a parameter defining the amount of the positive-energy
phase space taken into account. At the same time, since all hole-like
quasiparticle states, $N_n$$<$1/2, have negative values of $\bar{e} _{n}$
(\ref{equi3}), condition (\ref{cutoff}) quarantees that they are all taken into
account. In this way, we have a global cut-off prescription independent of
$\Omega^\pi$, which fulfills the requirement of taking into account the
positive-energy phase space as well as all quasiparticle states up to the highest
hole-like quasiparticle energy.

\subsection{Two-neutron separation energies and Fermi energies}

\label{sec3c}

A particular thrust of our analysis will be to identify the location of the
two-neutron drip line. The self-consistent HFB variational procedure produces two
quantities that provide information of relevance. One is the two-neutron
separation energy, $S_{2n}$, defined as the difference between the HFB energy for
the $N$$-2$ and $N$ neutron systems (with the same proton number) and the other
is the Fermi energy, $\lambda_{n}$.

The two-neutron separation energy provides ``global'' information on the total
$Q$-value corresponding to a hypothetical simultaneous transfer of two neutrons
into the $N$$-$2 ground state, leading to the ground state of the nucleus with
$N$ neutrons. The $Q$-value includes information on all differences in the
ground-state properties of both nuclei, like pairing, deformation, configuration,
etc. Whenever this $Q$-value becomes negative, the window for the spontaneous and
simultaneous emission of two neutrons opens up, and the nucleus with $N$ neutrons
is formally beyond the two-neutron drip line.

The Fermi energy, on the other hand, gives ``local'' information on the stability
of the given nucleus at a given pairing intensity, deformation, and
configuration. Within the HFB theory, the sign of the Fermi energy dictates the
localization properties of the HFB wave function; it is localized if
$\lambda_{n}$$<$0 and unlocalized (i.e., behaves asymptotically as a plane wave)
if $\lambda_{n}$$>$0. Thus, within the HFB approach, nuclei with
$\lambda_{n}$$>$0 spontaneously emit neutrons, while those with $\lambda_{n}$$<$0
do not emit neutrons, irrespective of the available $Q$-values for the real
emission. As such, we must take into account all solutions with $\lambda_{n}$$<$0
in discussing our self-consistent HFB results.

We will indeed see examples in Sec.~\ref{sec4} in which the nucleus has a
negative two-neutron separation energy, so that it is formally beyond the
two-neutron drip line, but nevertheless is localized and does not spontaneously
spill off neutrons.

\section{Results}

\label{sec4}

In this section, we present the results of several sets of HFB calculations
performed in the axial-deformed THO basis. All the calculations were carried out
using the Skyrme interaction SLy4 \cite{[Cha98]}, which has recently been
adjusted to the properties of stable nuclei, neutron-rich nuclei and neutron
matter. This force has a proven record in deformed-mean-field calculations \cite
{[Li96],[Cwi96],[Rut97],[Ter97a],[Hee98],[Nau98],[Sat98a],[Bur98]}, including
calculations of rotational properties in nuclei \cite
{[Rud98],[Bou98],[Bou98a],[Rud98a],[Rig99]}. At the same time, it reproduces the
masses of spherical nuclei with an accuracy similar to several other Skyrme
forces.

Below we review our choice of the various parameters that define our calculations
and present several tests of the THO approach. Then we present results obtained
for the Mg isotopes and for light nuclei at the two-neutron drip line. A more
extensive set of calculations will be presented in a future study, where we shall
also explore in detail the influence of the type of pairing force on the
properties of drip-line nuclei.

\subsection{Parameters and numerical details of the calculation}

\label{sec4a}

In all of the calculations reported here, we use a contact interaction
(\ref{delta}) in the p-p channel, which leads to volume pairing correlations
\cite{[Dob96]}. Following the discussion of Sec.~\ref{sec3b}, the pairing phase
space has been defined by a cut-off energy (see Eq.~(\ref{cutoff})) of
$\bar{e}_{\max}$=30\,MeV. This constitutes a very safe limit, for which all
convergence properties are well satisfied (see the discussion in
Refs.~\cite{[Dob96],[Ben99a]}). Within this phase space, the pairing strength
$V_{0}$ (see Eq.~(\ref{delta})) has been adjusted in a manner  analogous to the
prescription used in Ref.~\cite{[Dob95c]}, namely so that the average neutron
pairing gap (\ref{avdel}) for $^{120}$Sn equals the experimental value of $\Delta
_{n}$=1.245\,MeV. The resulting value is $V_{0}$=$-206$\,MeV\,fm$^3$. As
demonstrated in the Appendix of Ref.~\cite{[Dob96]}, changes in the cut-off
parameter $\bar{e}_{\max}$, leading to a renormalization of the pairing strength
$V_{0}$, can be safely disregarded when compared to all other uncertainties in
the methods used to extrapolate to unknown nuclei.

Although our axially-deformed HFB+THO code is able to work with arbitrary axial
oscillator lengths $L_\rho$ and $L_z$, we have used in these calculations a
spherical basis defined by a single common oscillator length $L_0$ (\ref
{lengthax}) (see Sec.~\ref{sec2c}). When optimizing the THO basis parameters
$L_0$, $a$, and $c$ (to minimize the total energy), we invariably find that for
weakly-bound nuclei the resulting exponential decay constant (\ref{dc}) is very
close to that given by the HFB estimate (\ref{hfbdc}). Based on this observation,
we have chosen to eliminate the THO parameter $a$ and to fix it in such a way
that the basis decay constant (\ref{dc}),  at the self-consistent solution, is
equal to the HFB decay constant (\ref{hfbdc}). In this way, we only have two
variational parameters in our calculations, $L_0$ and $c$.  The minimizations
were carried out independently for each nucleus. When describing each specific
application, we will indicate the number of shells included both in the
minimization that determines the LST parameters ($N^{\text{par}}_{\text{sh}}$)
and in the final calculations ($N_{\text{sh}}$).

All Gauss integrations were performed with 22 nodes in the $\rho$ direction and
24 nodes in the $z$ direction (due to the reflection symmetry assumed with
respect to the $x$-$y$ plane, only 12 nodes for $z$$>$0 were effectively
needed).

\subsection{Tests of the method}

\label{sec4b}

As the first test of the method, we considered doubly-magic nuclei. Such nuclei
are known to be spherical and thus amenable to reliable calculation using the
coordinate-space HFB code \cite{[Dob84]}.  By studying the extent to which our
code is able to reproduce the coordinate-space results (referred to in subsequent
discussion as {\em exact}), we can assess the method.  [We should note here that
HFB in fact reduces to HF in doubly-magic nuclei, since all pairing correlations
vanish].

These calculations were carried out both for nuclei along the beta-stability line
and for the very neutron-rich nucleus $^{28}$O. Some discussion of this latter
nucleus is in order here. $^{28}$O is known experimentally to be unbound
\cite{[Tar97],[Sak99]}, but is predicted to be bound in most mean-field
calculations \cite{[Kru97]}. Due to rapid changes of the single-particle energies
with neutron number, shell-model calculations \cite{[Cau98]} are able to explain
the sudden decrease of separation energies that occurs in the chain of oxygen
isotopes and renders $^{26}$O and $^{28}$O unbound. This effect seems to require
modifications to the effective interactions currently in use in mean-field
studies of light nuclei. Nevertheless, it is common to use $^{28}$O as a testing
ground of mean-field calculations near the neutron drip line, because according
to the standard magic-number sequence it is doubly magic and because it is
located (in typical mean-field calculations) just before the two-neutron drip
line. This is the philosphy underlying our inclusion of $^{28}$O. For comparison,
the configurational calculations were performed both in the HO and THO bases. To
assess the convergence of the results in the two cases, we varied the number of
HO major shells included, considering  $N_{\text{sh}}$=8, 12, 16, and 20. For a
given number of the major shells, we  minimized the total HF energies with
respect to the basis parameters, $L_{0}$ for the HO basis, and $L_{0}$ and $c$
for the THO basis,
so here $N^{\text{par}}_{\text{sh}}$=$N_{\text{sh}}$.
We also tested our HO axial-basis results obtained at any
given $N_{\text{sh}}$ with those available from Cartesian-basis calculations
\cite{[Dob97]} and the results agreed perfectly. Lastly, for the THO basis, we
compared with the calculations of Ref.~\cite{[Sto98b]}, where spherical symmetry
was imposed, and obtained identical results.

\begin{figure}[tbh]
\begin{center}
\leavevmode
\epsfig{file=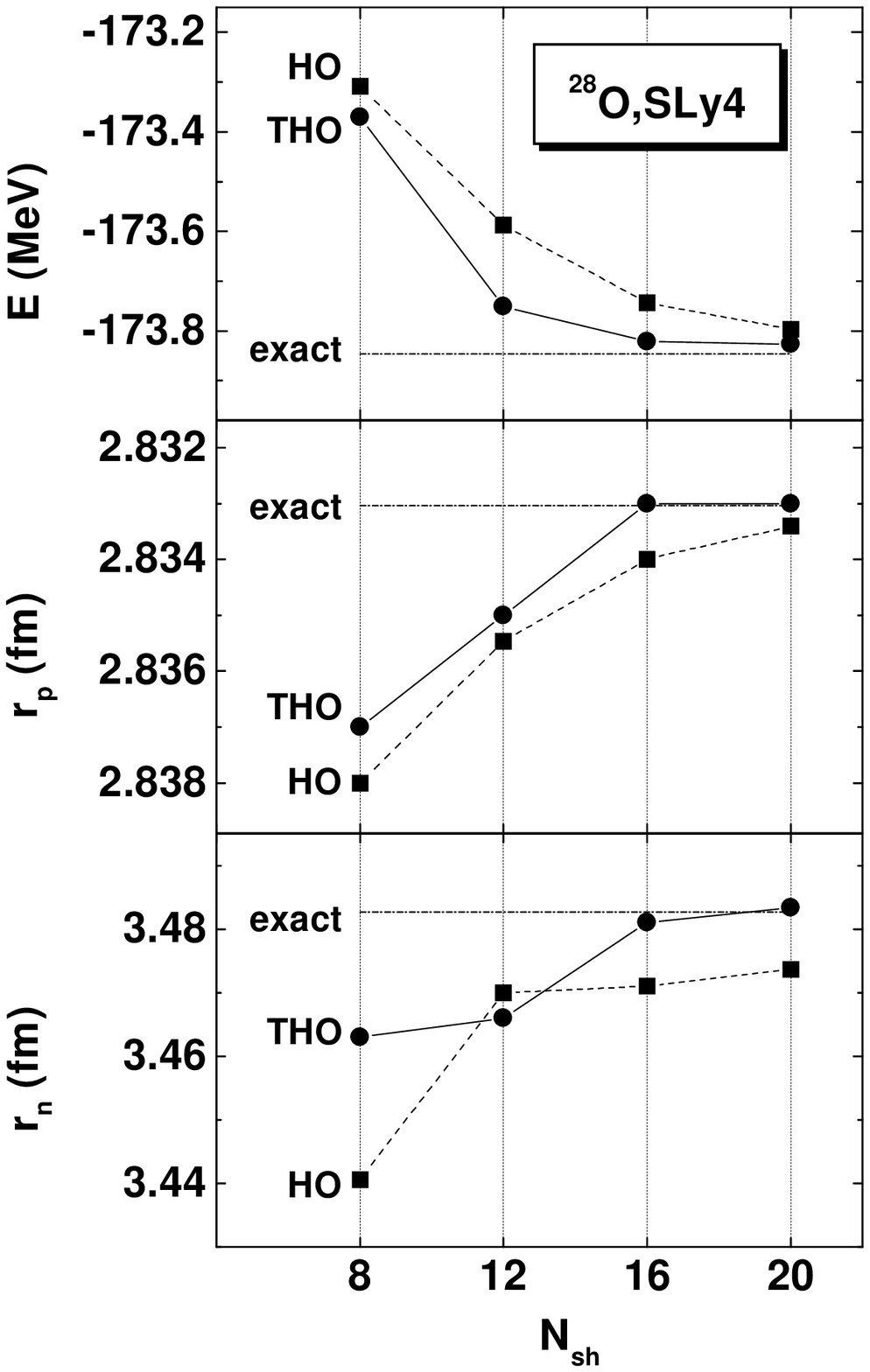, width=12cm}
\end{center}
\caption[C1]{Total energies $E$, and proton and neutron rms radii, $r_p$ and
$r_n$, obtained in the HFB+SLy4 calculations for $^{28}$O by using the HO
and THO bases, as functions of the number of HO shells $N_{\text{sh}}$.
The exact results refer to those obtained from spherical coordinate-space
calculations. }
\label{fig01}
\end{figure}

As expected, for nuclei within the valley of beta stability the HO and THO
results are close to one another and, furthermore, coincide with the exact HFB
(HF) results. The situation is quite different for the neutron-rich nucleus
$^{28}$O, for which the calculations indicate the presence of a significant
neutron skin. In Fig.~\ref{fig01}, we present the HO and THO results for the
total energy and for the proton and neutron rms radii as functions of
$N_{\text{sh}}$. For each of the calculated observables, the exact results are
shown as a straight line as a function of $N_{\text{sh}}$. Clearly, when we
increase the number of major shells, both the HO and THO results for the total
energy and for the proton rms radius converge to the exact HFB values. In
contrast, the HO neutron rms radius still differs from the exact value, even at
$N_{\text{sh}}$=20, while the THO basis gives the correct result.

\begin{figure}[tbh]
\begin{center}
\leavevmode
\epsfig{file=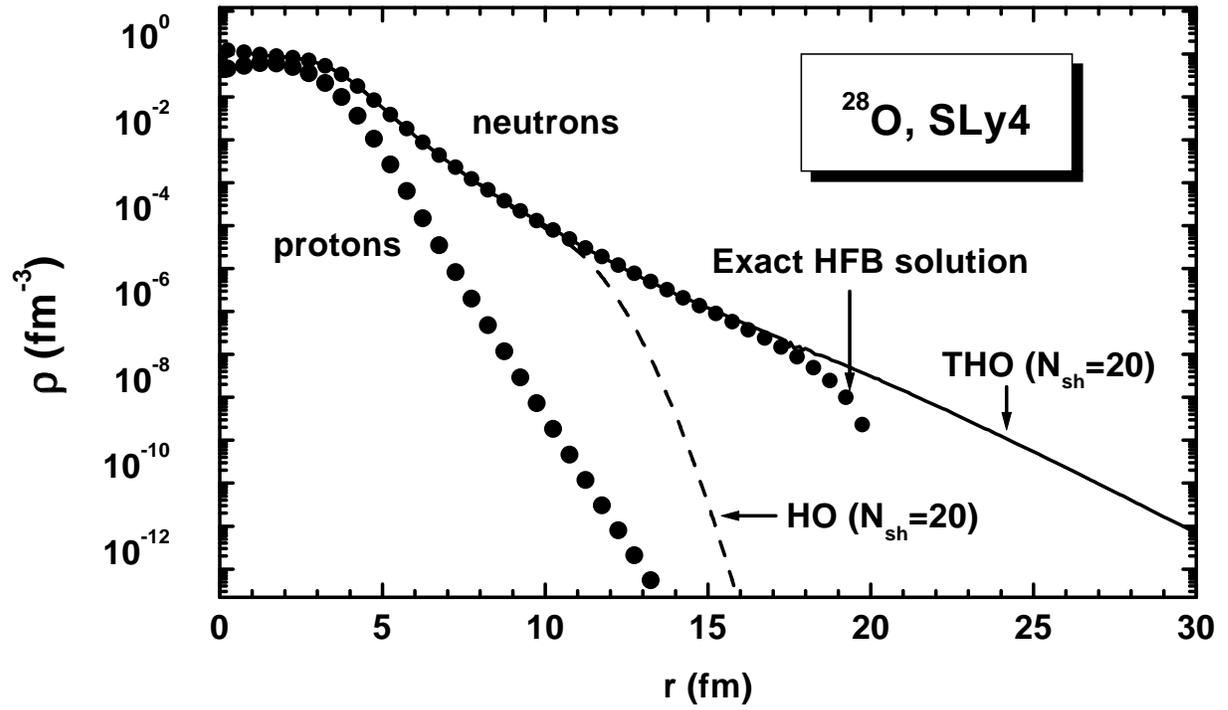, width=16cm}
\end{center}
\caption[C1]{Neutron densities obtained in the HFB+SLy4 calculations for
$^{28}$O by using the HO (dashed line) and THO (solid line) bases. Neutron
and proton densities denoted as ``exact'' (dots) have been obtained from
spherical coordinate-space calculations in a box of $R_{\text{box}}$
=20\,fm. }
\label{fig02}
\end{figure}

An explanation of this difference becomes clear when looking at Fig.~\ref
{fig02}, in which we compare (in logarithmic scale) the HO and THO neutron
densities with those from the exact HFB calculations. The HO neutron density
fails to reproduce the correct asymptotic behavior at large distances (see also
the discussion in Ref.~\cite{[Dob96]}). The THO density, on the other hand, shows
perfect agreement with the exact HFB density. There is a difference,  of course,
near and beyond the box boundary ($R_{\text{box}}$=20 fm is used in the
coordinate HFB calculations).  The coordinate-space density rapidly falls to zero
at the boundary, while the THO density continues with the correct exponential
shape out to infinite distances.

It is clear that the rather small numerical discrepancy between the HO and THO
neutron rms radii (Fig.~\ref{fig01}) does not reflect the seriousness of the
error in neutron densities that arises when using the HO basis. It is also
obvious that observables which do not strongly depend on neutron densities at
large distances, like the total energy or proton radii, are fairly well
reproduced in standard HO calculations. On the other hand, observables that do
depend on densities in the outer region, most notably pairing correlations
\cite{[Dob96]}, require the correct asymptotic behavior provided by the THO
basis.

Encouraged by the excellent results in spherical nuclei, where a comparison with
reliable coordinate-space calculations was possible, we next turned to deformed
systems. Here, since no coordinate-space HFB results are available, our tests
were limited to a study of the convergence of results with increasing number of
HO shells. [The exact results would be obtained in either the HO or THO expansion
with a complete space, i.e., an infinite number of shells.] Whenever the number
of HO shells used in the final HFB calculation was 12 or less, we determined the
basis parameters with that same number of shells,
$N^{\text{par}}_{\text{sh}}$=$N_{\text{sh}}$. When the number of HO shells of
the final calculations exceeded 12, however, we still determined the basis
parameters with $N^{\text{par}}_{\text{sh}}=12$.

\begin{figure}[tbh]
\begin{center}
\leavevmode
\epsfig{file=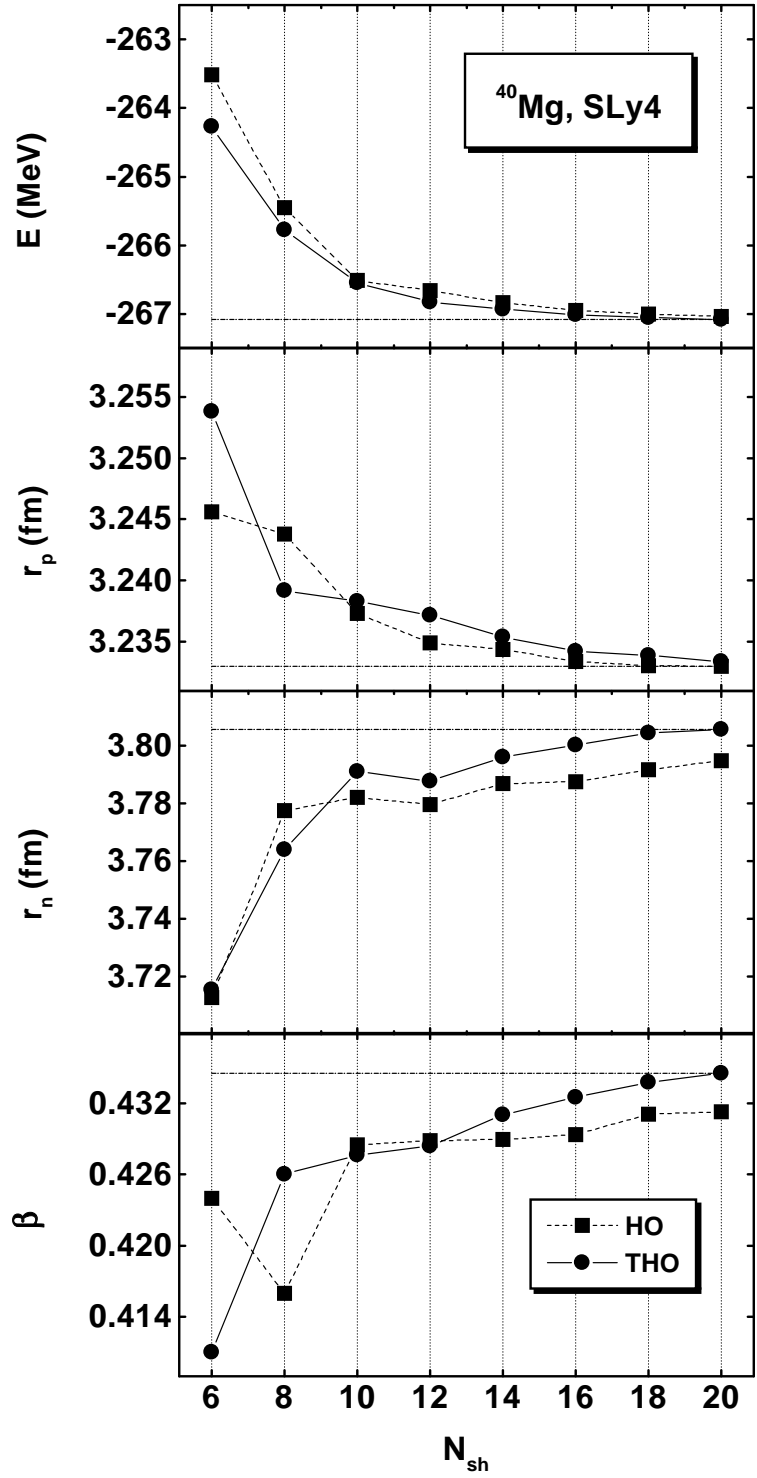, width=10cm}
\end{center}
\caption[C1]{Total energies $E$, proton and neutron rms radii, $r_p$ and $r_n
$, and deformations $\beta$ obtained in the HFB+SLy4 calculations for 
$^{40}$Mg by using the HO and THO bases, as functions of the number of HO shells 
$N_{\text{sh}}$. The horizontal lines denote the THO results obtained at 
$N_{\text{sh}}$=20. }
\label{fig03}
\end{figure}

\begin{figure}[tbh]
\begin{center}
\leavevmode
\epsfig{file=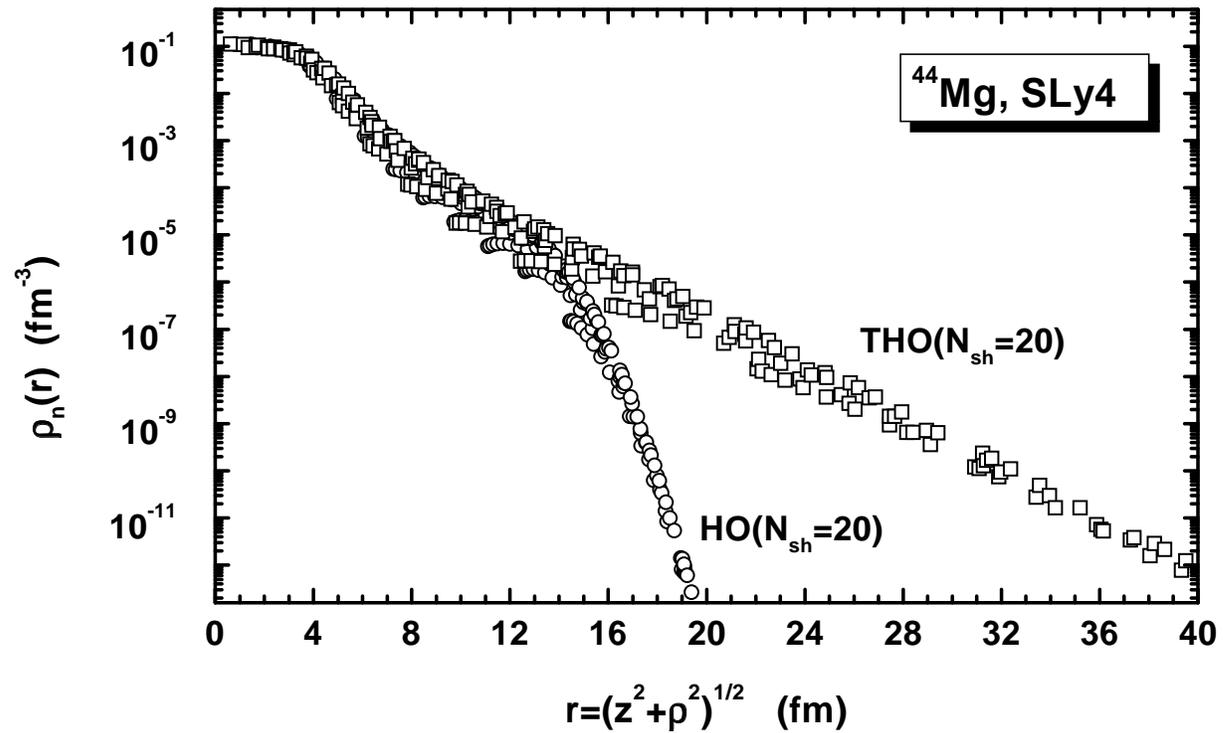, width=16cm}
\end{center}
\caption[C1]{Neutron densities obtained in the HFB+SLy4 calculations for the
deformed ground state of $^{44}$Mg by using the HO (circles) and THO
(squares) bases. Each point corresponds to one Gauss-integration node in the
$z$-$\rho$ plane, and the results are plotted as functions of the distance
from the origin, $r$=$(z^2$+$\rho^2)^{1/2}$. }
\label{fig04}
\end{figure}

In Fig.~\ref{fig03}, we show convergence results for the ground state of the
weakly-bound deformed nucleus $^{40}$Mg. The top three panels give the results
for the total energies, the proton rms radii and the neutron rms radii,
respectively. The fourth gives results for the $\beta$ deformation, which is
related to the quadrupole moment $\langle Q \rangle$ ($Q= \sum_{i=1}^A 2z_i^2-x_i^2-y_i^2$) 
and the rms
radius $\langle r^2\rangle$ by
\begin{equation}
\beta =\sqrt{\frac{\pi}{5}}\frac{\langle Q\rangle}{A\langle r^2\rangle}.
\label{beta}
\end{equation}
The results obtained with $N_{\text{sh}}$=20  are indicated in the figure by horizontal lines.
Again, both bases yield very good convergence for the total energy and proton
radius. In contrast, noticeable differences between the HO and THO results can be
seen for the deformation and neutron rms radius, and they persist to large values
of $N_{\text{sh}}$. Although these differences are small in magnitude, they are
caused by a very large error in the HO neutron density distribution. This is
illustrated in Fig.~\ref{fig04}, where we show the neutron densities calculated
for the nearby $^{44}$Mg nucleus. Every point in
the figure corresponds to the value of the
neutron density at a given Gauss-integration node. Since there are always several
nodes near a sphere of the same radius $r=\sqrt{z^{2}+\varrho^{2}}$, there can be
some scatter of points, corresponding to different densities in different
directions. This is especially true at small distances. At large distances, the
scatter is greatly reduced and the densities exhibit to a good approximation
spherical asymptotic behavior, exponential in the case of the THO expansion and
Gaussian in the case of the HO expansion. Note, however, that some scatter
persists in the THO results out to large distances, suggesting that deformation
effects are still present there. This is apparently reflecting the importance of
deformation of the least-bound orbitals. Clearly, the asymptotic properties of
the HO and THO neutron densities are very different from one another, as they
were in the spherical calculations (see Fig.~\ref{fig02}).

\subsection{Drip-line-to-drip-line calculations in Mg}

\label{sec4c}

The chain of even-$Z$ magnesium isotopes has been the subject of numerous recent
theoretical analyses. The extreme interest in this isotopic chain is motivated by
recent measurements in $^{32}$Mg \cite {[Mot95],[Hab97],[Aza98]}, which show a
larger-then-expected quadrupole collectivity. Based on the relativistic and
non-relativistic mean-field approaches and on shell-model calculations (see
Ref.~\cite{[Rei99]} for a review), it is now well documented that shape
coexistence and configuration mixing occur in this $N$=20 nucleus. Moreover,
recent advances in radioactive-ion-beam technology allow mass measurements of
even heavier isotopes \cite{[Sar99]}, giving hope that the neutron-drip line can
be experimentally reached in the $Z$=12 chain \cite{[Tan98a]}.

In this section, we present results of an investigation of the deformation
properties of the even-even Mg isotopes from the proton-drip-line to the
neutron-drip-line. Our results are complementary to those of recent Skyrme+HFB
calculations \cite {[Ter97a]}, in which the imaginary-time evolution method of
finding eigenstates of the mean-field Hamiltonian $h$ (see Sec.~\ref{sec4}) was
combined with a diagonalization of the HFB Hamiltonian within a relatively small
set of these eigenstates. In that study, a complete set of results was given only
for the SIII force and density-dependent pairing was used. Here, we present a
complete set of results for the SLy4 force with a density-independent (volume)
pairing interaction. These calculations were carried out with $N_{\text{sh}}$=20
and $N^{\text{par}}_{\text{sh}}$=12 HO shells.

\begin{figure}[tbh]
\begin{center}
\leavevmode
\epsfig{file=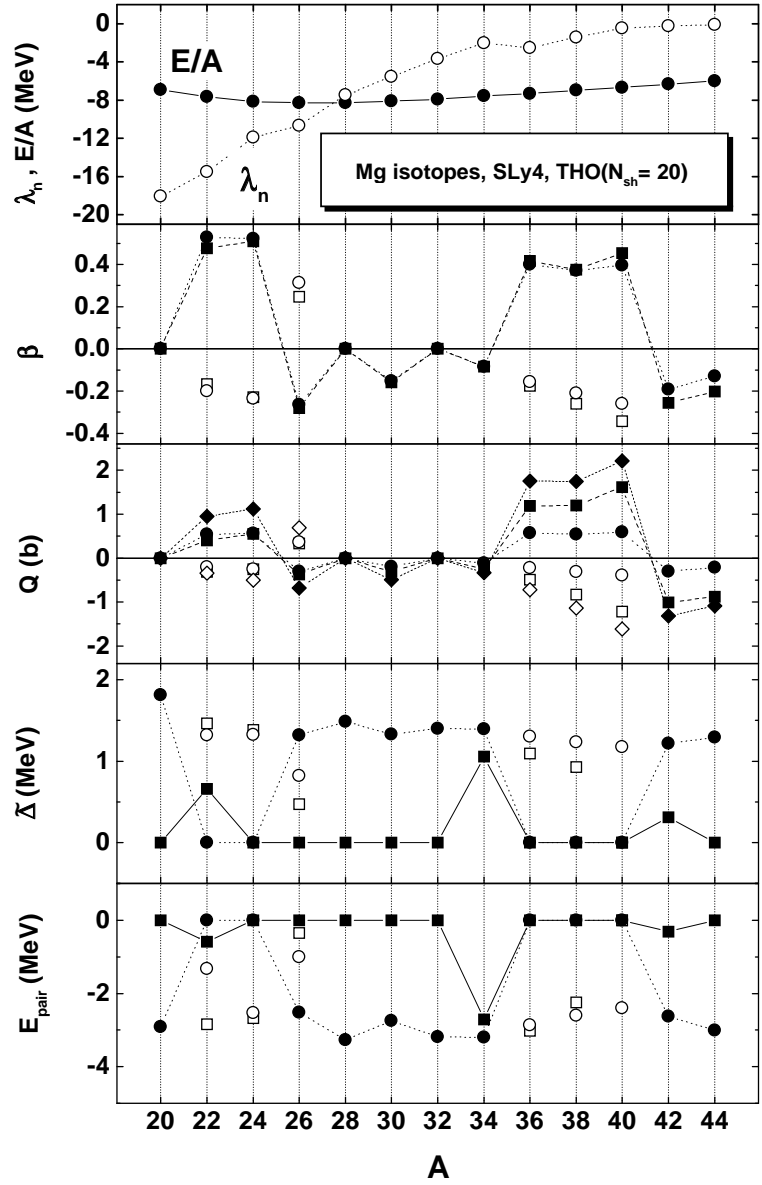, width=10cm}
\end{center}
\caption[C1]{Neutron Fermi energies $\lambda_n$, energies per particle $E/A$,
deformations $\beta$, quadrupole moments $Q$, pairing gaps $\widetilde
\Delta$, and pairing energies $E_{\text{pair}}$ calculated for the Mg
isotopes within the HFB+SLy4 method in the THO basis ($N_{\text{sh}}$=20),
as functions of the mass number $A$. Apart from the upper panel, circles,
squares, and diamonds pertain to proton, neutron, and total results,
respectively. Closed symbols connected with lines denote values for the
absolute minima in the deformation-energy curve (axial shapes are assumed),
while open symbols pertain to secondary minima. }
\label{fig05}
\end{figure}

\begin{figure}[tbh]
\begin{center}
\leavevmode
\epsfig{file=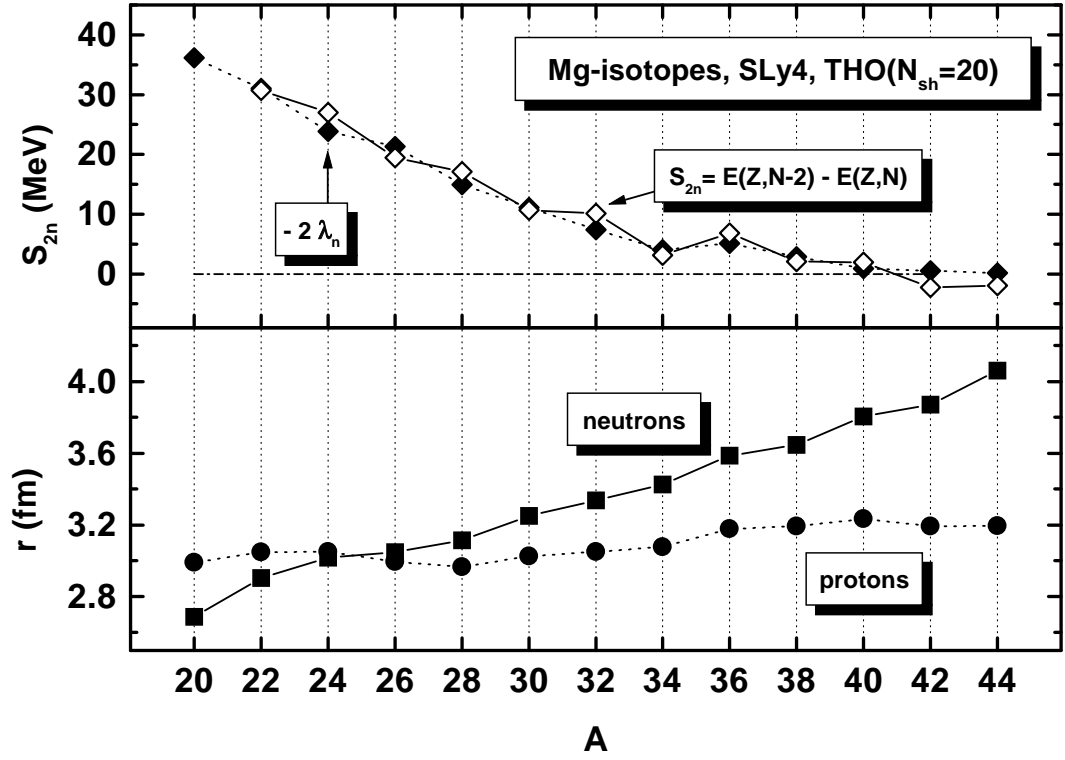, width=14cm}
\end{center}
\caption[C1]{Upper panel: two-neutron separation energies $S_{2n}$ (open
symbols) compared to $-2\lambda_n$ (closed symbols), and lower panel: proton
and neutron rms radii. Calculations for the Mg isotopes were performed
within the HFB+SLy4 method in the THO basis for $N_{\text{sh}}$=20. }
\label{fig06}
\end{figure}

In Fig.~\ref{fig05}, we plot the total HFB energies per nucleon $E/A$, the
neutron chemical potentials $\lambda _{n}$, the neutron and proton deformation
parameters, $\beta_{n}$ and $\beta_{p}$, the neutron, proton, and total
quadrupole moments, $Q_{n}$, $Q_{p}$, and $Q_{t}$, the average neutron and proton
pairing gaps, $\widetilde{\Delta}_{n}$ and $\widetilde{\Delta}_{p}$,
Eq.~(\ref{avdel}), and the pairing energies $E_{\text{pair}}^{n}$ and
$E_{\text{pair}}^{p}$ for the magnesium isotopes as functions of the mass number $A$.
Ground-state values are shown by full symbols connected by lines, while the
isolated open symbols correspond to secondary minima of the deformation-energy
curves. In the top panel
of Fig.~\ref{fig06}, we compare the results for the two-neutron separation
energies $S_{2n}$ (open symbols) with those for the related quantity
$-2\lambda_{n}$ (full symbols), and in the bottom panel we show the neutron and
proton rms radii.

The lightest Mg isotope predicted by these calculations to be bound against
two-proton decay is $^{20}$Mg.  The heaviest bound against two-neutron decay, on
the basis of having a positive two-neutron separation energy, is $^{40}$Mg. On
this basis, the position of the two-neutron drip line obtained within the
HFB+SLy4 approach is identical to that obtained in the finite-range droplet model
\cite{[Mol95]}, relativistic mean field (RMF) \cite{[Ren96]}, and HFB+SIII
\cite{[Ter97a]} calculations. The RMF approach with the NL-SH effective
interaction \cite{[Lal98]} predicts the two-neutron drip line at $^{42}$Mg, and
the relativistic Hartree-Bogoliubov (HB) approach with the NL3 effective
interaction \cite {[Lal98a]} predicts it at or beyond $^{44}$Mg.

On the other hand, from Fig.~\ref{fig06}, we see that both $^{42}$Mg and
$^{44}$Mg, though having negative values of $S_{2n}$, have
(small) negative values of the
Fermi energy, $\lambda_{n}$. According to the discussion of Sec.~\ref{sec3b},
these nuclei, both of which exhibit oblate shapes, are bound against neutron
emission. We will return to this point later.

The most deformed nucleus of the isotope chain is $^{24}$Mg with almost the same
neutron and proton deformations. At the other end of the chain, due to a large
excess of neutrons over protons, significant differences exist between the proton
and neutron quadrupole moments. The onset of large deformation in $^{36}$Mg
causes a decrease of the neutron chemical potential $\lambda_{n}$ with respect to
its value in $^{34}$Mg. This gives an additional binding of $^{36}$Mg, and
correspondly to an increase and decrease of the two-neutron separation energies
$S_{2n}$ in $^{36}$Mg and $^{38}$Mg, respectively (see Fig.~\ref{fig06}). In
experiment \cite{[Sar99]}, these changes are less pronounced and arrive two mass
units earlier, giving rise to a small and large decrease of $S_{2n}$ in $^{34}$Mg
and $^{36}$Mg, respectively.

Concerning the ground-state deformation properties (full symbols connected by
lines in Fig.~\ref{fig05}), the proton drip-line nucleus $^{20}$Mg displays a
well defined spherical minimum ($N$=8 is a magic number). Then, there is a
competition between prolate ($^{22,24}$Mg and $^{36,38,40}$Mg) and oblate
($^{26,30}$Mg) deformations, while $^{28,32}$Mg are spherical. The last two
localized isotopes (with negative Fermi energies), $^{42,44}$Mg, display oblate
deformations. Secondary minima of the deformation energy curves (isolated
symbols) exist for isotopes $^{22,24,26}$Mg and $^{36,38,40}$Mg.

Non-zero proton pairing correlations are present at all spherical or oblate
minima. However, at these shapes, tangible neutron pairing exist only in
$^{22,24}$Mg and $^{34,36,38}$Mg. Moreover, for all nuclei with prolate
ground-state shapes, i.e., in $^{22,24}$Mg and $^{36,38,40}$Mg, both proton and
neutron correlations are small or vanish altogether. These results are at
variance with the Gogny-pairing HB calculations of Ref.~\cite{[Lal98a]}, where
non-zero pairing exists in all the heavy Mg isotopes. Also, in Ref.~\cite
{[Ter97a]}, stronger pairing correlations were obtained for the zero-range
density-dependent pairing force. However, in that study, the strength parameters
were not adjusted to odd-even mass staggering but rather taken from high-spin
calculations of superdeformed bands. Our results suggest that the pure HFB-pairing
approach is not necessarily the best way to treat pairing correlations in the Mg
isotopes, and approximate or exact particle-number projection should probably be
employed.

At this point, it is worth expanding a bit on the unusual results for
$^{42,44}$Mg. In these two isotopes, the solutions corresponding to prolate
shapes are unstable ($\lambda_{n}$$>$0), while those corresponding to oblate
shapes continue to be bound, i.e., they have $\lambda_{n}$=$-$0.253 and
$-$0.092\,MeV for $A$=42 and 44, respectively. The bound ground states of these
two nuclei are thus oblate, whereas in the lighter isotopes the oblate solutions
corresponded to secondary minima. This is the origin of the sudden change in 
two-neutron separation energies, which become negative in $^{42}$Mg and $^{44}$Mg 
($S_{2n}$=$-$2.237 and $-$1.975\,MeV, respectively. In the case of
$^{42}$Mg, however, two-neutron emission should be hindered by the fact that the
parent and daughter nuclei have dramatically different shapes, and, by this token,
$^{42}$Mg may still have a substantial half-life even though it is beyond the
two-neutron drip line.

The bottom panel of Fig.~\ref{fig06} shows the neutron and proton rms radii,
$r_{n}$ and $r_{p}$. At the proton drip-line, the neutron rms radii are smaller
than the proton rms radii, and then they increase with increasing neutron number.
Around $^{24,26}$Mg, $r_{n}$ becomes almost equal to $r_{p}$, and for nuclei close
to the neutron drip-line, $r_{n}$ takes significantly larger values than $r_{p}$.
The increase of $r_{n}$ is fairly linear, similarly as in
Refs.~\cite{[Ter97a],[Lal98],[Lal98a]}, and gives no hint of an existence of
unusually larger neutron distributions at the neutron drip line (see also
the discussion in Refs.~\cite{[Dob99a],[Miz99]}).

\subsection{Neutron-drip-line calculations}

\label{sec4d}

Having at our disposal a viable method for performing deformed HFB calculations
up to the drip lines, we have performed a systematic study of the equilibrium
properties of the neutron-rich nuclei in all even-$Z$ isotopic chains with proton
numbers from $Z$=2--18. In this way, we have explored the
neighborhood of the neutron-drip line for all neutron numbers from $N$=6--40.

We
first performed spherical HFB+SLy4 calculations in coordinate space, using the
methods and the code developed in Ref.~\cite{[Dob84]}. We used volume delta
pairing, with a coupling constant $V_0$=$-$218.5\,MeV\,fm$^3$, adjusted as in
Ref.~ \cite{[Dob95c]}. This value is very close to the one used in our deformed
THO code (see Sec.~\ref{sec4a}), suggesting that the effective pairing phase
spaces used in the two approaches are very similar to one another.

{}From the spherical calculations, we obtained that the heaviest even isotopes,
for which the Fermi energies are negative are: $^{8}$He, $^{12}$B, $^{22}$C,
$^{28}$O, $^{30}$Ne, $^{44}$Mg, $^{46}$Si, $^{50}$S, $^{58}$Ar.  We used these
spherical results as a starting point for our deformed calculations.

\begin{figure}[tbh]
\begin{center}
\leavevmode
\epsfig{file=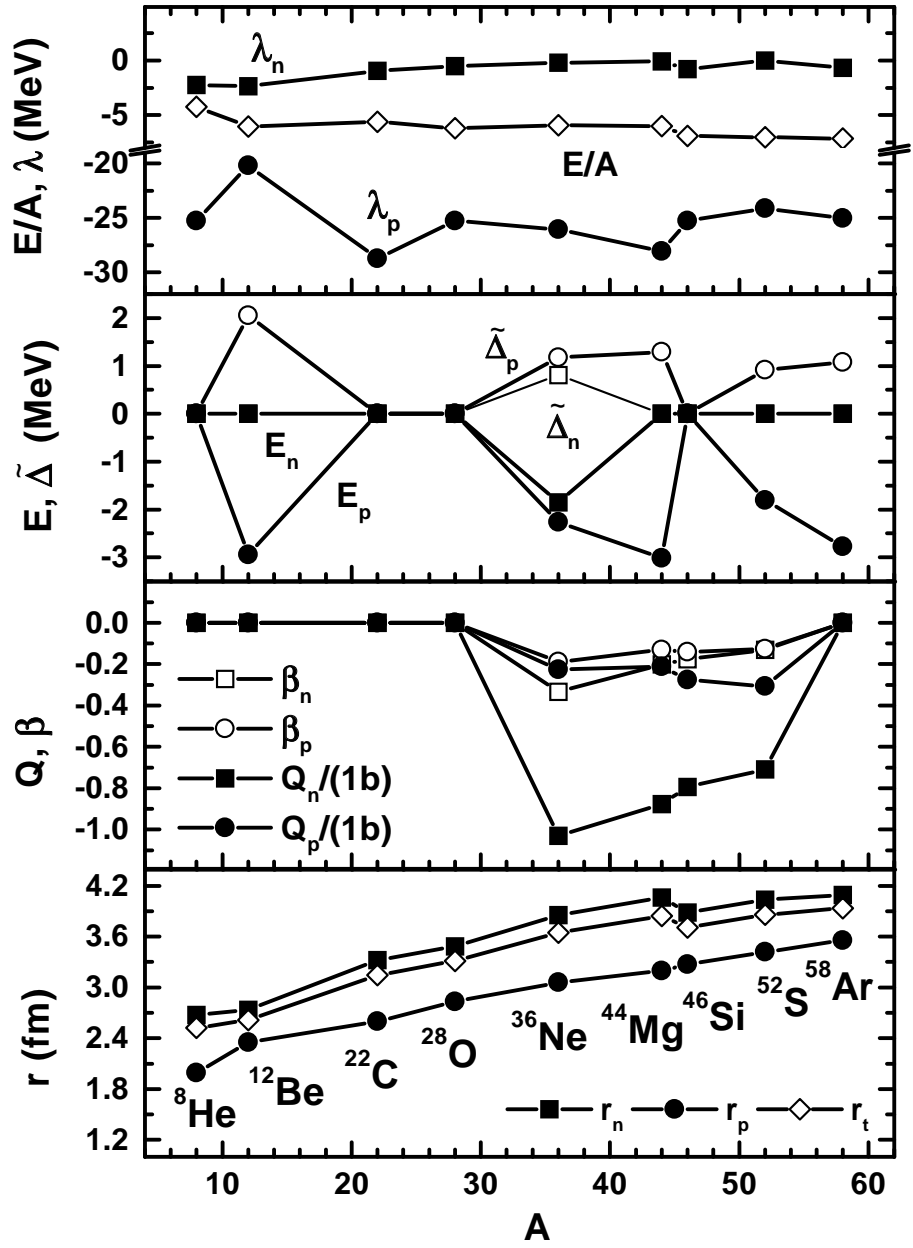, width=12cm}
\end{center}
\caption[C1]{Neutron Fermi energies $\lambda_n$, energies per particle $E/A$,
pairing gaps $\widetilde\Delta$, pairing energies $E$, deformations $\beta$,
quadrupole moments $Q$, and rms radii $r$ calculated for the
neutron-drip-line nuclei (indicated in the lower panel) within the HFB+SLy4
method in the THO basis, as functions of the mass number $A$. Circles,
squares, and diamonds pertain to proton, neutron, and total results,
respectively. }
\label{fig07}
\end{figure}

Next, within the deformed THO formalism, we found that the heaviest isotopes with
negative Fermi energies are: $^{8}$He, $^{12}$B, $^{22}$C, $^{28}$O, $^{36}$Ne,
$^{44}$Mg, $^{46}$Si, $^{52}$S, $^{58}$Ar. The results obtained for these nuclei
are summarized in Fig.~\ref{fig07}. By comparing the deformed results to the
spherical results, we see that the position of the last bound nucleus is
influenced by deformation only in $^{36}$Ne and $^{52}$S. Volume pairing
correlations are very weak in these nuclei; indeed, in all but the one case of
$^{36}$Ne, neutron pairing vanishes in the last bound nucleus of an isotopic chain. 
This suggests the
necessity of using a surface pairing force here. Such a conclusion is supported
by the fact that HB calculations \cite{[Lal98a]}, carried out with a Gogny
pairing force, give sizable neutron pairing correlations in this region (Note
that surface pairing and Gogny pairing produce quite similar distributions
\cite{[Dob96]} over the single-particle states.)

Since neutron pairing vanishes in $^{12}$B, our result is identical to that of
Ref.~\cite{[Li96]}, namely that the SLy4 force does not produce $^{14}$B as
bound, in disagreement with experiment \cite{[Aud93]}. Similarly, neither pairing
nor deformation effects are present in the calculated $^{28}$O nucleus, and hence
this nucleus remains bound (see discussion in Sec.~\ref{sec4b}). On the other
hand, the SLy4 force correctly describes $^{8}$He
\cite{[Aud93]} and $^{22}$C as the last bound nuclei of their respective
isotope chains \cite{[Yon98]}.

A remarkable result obtained in our calculations is that the last bound nuclei
for all chains of isotopes heavier than oxygen have oblate ground-state shapes.
In all of them, the mechanism for this effect is identical to that discussed for
the Mg isotopes (see Sec.~\ref{sec4c}), namely that the neutron Fermi energy
$\lambda_n$, as a function of the neutron number $N$, becomes positive for
smaller values of $N$ in the prolate ground states than it does in the oblate
secondary minima. Therefore, in the heaviest bound isotopes, the prolate states
are unbound, whereas the oblate states continue to be bound and become the
ground-state configurations.

\section{Conclusions}

\label{sec5}

In this paper, we have applied a local-scaling point transformation to the
deformed three-dimensional Cartesian harmonic oscillator wave functions so as to
allow for a modification of their unphysical asymptotic properties. In this way,
we have obtained single-particle bases that remain infinite, discrete, and
complete, but for which the wave functions have the asymptotic properties that
are required by the canonical bases of Hartree-Fock-Bogoliubov theory. These
bases preserve all the simplicity of the original harmonic-oscillator wave
functions, and at the same time are amenable to very efficient numerical methods,
such as Gauss-integration quadratures. They also allow for very simple
calculations of local densities, which are at the core of self-consistent methods
based on a Skyrme effective interaction.

The axial transformed harmonic oscillator basis has been implemented to achieve a
fast and reliable method for solving the HFB equations with the correct
asymptotic conditions. We have discussed several practical aspects of the
implementation, like the treatment of pairing correlations, and tested the
convergence and accuracy.

The formalism was developed for a general deformed transformed harmonic
oscillator basis. Practical application within the configurational HFB formalism
suggested a simplification to a purely spherical basis, however, as had been used
in earlier calculations. We have nevertheless presented the general formalism
because of its possible use in other applications.

As a first application of this new methodology, we have carried out HFB
calculations using the SLy4 Skyrme force and a density-independent (volume)
pairing force. The calculations were performed for the chain of even-$Z$ Mg
isotopes and for the light even-$Z$ nuclei located near the two-neutron drip
line. We have presented results for binding energies, quadrupole moments, and for
the pairing properties of these nuclei.

Perhaps the most interesting outcome of our calculations is that nuclei that are
formally beyond the two-neutron drip line, i.e., those with negative two-neutron
separation energies, may have tangible half lives, provided (i) that they have
localized ground states (negative Fermi energies), and (ii) their ground-state
configurations are significantly different than those of the (daughter) nuclei
with two less neutrons. According to our calculations, precisely such a situation
occurs in the chains of isotopes  with $Z$=10, 12, 14, 16 and 18. In these
chains, the prolate configuration becomes unbound before (i.e., for a smaller
neutron number) than the oblate configuration. That change in the ground state
structure leads to negative two-neutron separation energies and thus to the
exotic conditions given above.

\acknowledgments

This work has been supported in part by the Bulgarian National Foundation for
Scientific Research under project $\Phi $-809, by the Polish Committee for
Scientific Research (KBN) under Contract No.~2~P03B~040~14, by a computational
grant from the Interdisciplinary Centre for Mathematical and Computational
Modeling (ICM) of the Warsaw University, and by the National Science Foundation
under grant \#s PHY-9600445, INT-9722810 and PHY-9970749.


\clearpage

\onecolumn

\end{document}